\documentclass[prl,reprint,amssymb]{revtex4-2}

\usepackage{xcolor}

\usepackage{bm}

\usepackage{amsmath}  
\usepackage{amsfonts}
\usepackage{graphicx}
\usepackage{hyperref}
\hypersetup{
	colorlinks=true,
	citecolor=blue,
	linkcolor=red,
	urlcolor = blue
}

\begin{document}
	
	\title{Origins of electromagnetic radiation from spintronic terahertz emitters: \\ A time-dependent density functional theory plus Jefimenko equations approach}
	
	
	\author{Ali Kefayati}
	\author{Branislav K. Nikoli\'c}
	\email{bnikolic@udel.edu}
	\affiliation{Department of Physics and Astronomy, University of Delaware, Newark, DE 19716, USA}
	

	\begin{abstract}
		Microscopic origins of charge currents and electromagnetic (EM) radiation generated by them in spintronic THz emitters---such as, femtosecond laser pulse-driven single magnetic layer or its heterostructures with a nonmagnetic layer hosting strong spin-orbit coupling (SOC)---remain poorly understood  despite nearly three decades since the discovery of ultrafast demagnetization. We introduce a first-principles method to compute these quantities, where the dynamics of charge and current densities is obtained from real-time time-dependent density functional theory (TDDFT), which are then fed into the Jefimenko equations for properly retarded electric and magnetic field solutions of the Maxwell equations. By Fourier transforming different time-dependent terms in the Jefimenko equations, we unravel that in 0.1--30 THz range the electric field of far-field EM radiation by Ni layer, chosen as an example, is {\em dominated by charge current pumped by demagnetization}, while often invoked magnetic dipole radiation from time-dependent magnetization of a single magnetic layer is a negligible effect.  Such overlooked case of charge current pumping by time-dependent quantum system, whose magnetization is shrinking while its vector does not rotate, does not require any spin-to-charge conversion via SOC effects. In Ni/Pt bilayer, EM radiation remains dominated by charge current within Ni layer, whose magnitude is larger than in the case of single Ni layer due to faster demagnetization, while often invoked spin-to-charge conversion within Pt layer provides additional but smaller contribution. By using Poynting vector and its flux, we also quantify efficiency of conversion of light into emitted EM radiation, as well as the  angular distribution of the latter.
	\end{abstract}
	
	\maketitle

	{\em Introduction.}---Ultrafast-light-driven magnetic materials and their heterostructures have attracted considerable attention from both fundamental~\cite{Beaurepaire1996, Gillmeister2020, Wang2017a} and applied perspectives~\cite{Leitenstorfer2023, Seifert2016}. From the fundamental viewpoint, a femtosecond (fs) laser pulse drives magnetic materials into a far-from-equilibrium state~\cite{Gillmeister2020, Suresh2023} whose coupled charge and spin dynamics can probe the strength of exchange interactions, as well as the interplay between many-body excitations and spin correlations~\cite{Gillmeister2020}. Such highly nonequilibrium states are also of great interest to applications as their magnetization can be reversed~\cite{Kimel2019}  on sub-ps timescale (in contrast to its reversal in near-equilibrium magnets that takes much longer $\sim 100$ ps time); or ferromagnet/normal-metal (FM/NM) bilayers, where NM layer hosts strong spin-orbit coupling (SOC), have opened new avenues~\cite{Seifert2016} for  table-top emitters of ultrabroadband (in the range 0.3--30 THz) electromagnetic (EM) radiation at terahertz (THz) frequencies~\cite{Leitenstorfer2023}.   
	
\begin{figure}
		\centering
		\includegraphics[width = \linewidth]{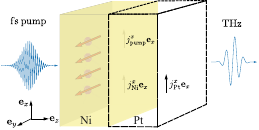}
		\caption{Schematic view of systems employed in  demagnetization~\cite{Beaurepaire1996,Rouzegar2022} and THz emission~\cite{Beaurepaire2004,Huang2019,Rouzegar2022,Seifert2016,Wu2017} experiments where fs laser pulse irradiates either a single FM layer (such as Ni we choose), or FM/NM bilayer where NM layer hosts strong SOC (such as Pt we choose). The thickness of Ni layer is 3 MLs, while in the second setup we add two MLs of Pt. The local magnetization (red arrows) of Ni layer is along the $y$-axis in equilibrium, as well as during demagnetization in nonequilibrium.}
		\label{fig:fig1}
	\end{figure}

	However, microscopic understanding of laser pulse-driven magnetization dynamics and the ensuing generation of currents and EM radiation is far from complete due to many competing mechanisms becoming relevant on different $\sim 10$ fs  time segments~\cite{Siegrist2019,Tengdin2018,Krieger2015}. Although first-principles simulations of realistic materials, based on real-time time-dependent density functional theory (TDDFT)~\cite{Krieger2015, Krieger2017, Dewhurst2018, Dewhurst2021, Chen2019a, Chen2019c, Mrudul2023, Simoni2017, Stamenova2016,Acharya2020,Elliott2022,Pellegrini2022,Simoni2022}, have provided a plethora of insights---such as, angular momentum transfer~\cite{Chen2019a,Chen2019c,Dewhurst2021,Simoni2022} between photons, electrons and ions; spin transport across interfaces~\cite{Elliott2022}; and spin-flips mediated by SOC~\cite{Krieger2015}---they have been largely focused on computing magnetization vs. time, $\mathbf{M}(t)$. Thus,  computation of spin and charge currents, as well as the ensuing EM radiation generated by the latter as {\em directly observed}  quantity~\cite{Rouzegar2022,Beaurepaire2004,Huang2019,Wu2017,Seifert2016,Zhang2020}, remain largely unexplored. This forces experiments to rely on plausible interpretations of  measured THz radiation, such as that: a single FM layer emits~\cite{Rouzegar2022,Beaurepaire2004,Huang2019} magnetic dipole radiation due to $\mathbf{M}(t)$; or that interlayer spin current in FM/NM bilayers [Fig.~\ref{fig:fig1}] must be efficiently converted~\cite{Seifert2016, Rouzegar2022, Wu2017} into parallel-to-the-interface charge current within NM layer whose time dependence is the main source of THz radiation. The spin-to-charge conversion mechanisms invoked include the inverse spin Hall effect (ISHE)~\cite{Saitoh2006} in the bulk of NM layer or other SOC-driven mechanisms at FM/NM interfaces~\cite{Jungfleisch2018a, Gueckstock2021, Wang2023}. The demand to compute EM radiation and understand which mechanisms  contribute the most to it is further emphasized by the fact that experiments analyze emitted THz signal to extract the temporal shape of charge, and from it spin, current, and then infer from the latter underlying ultrafast spin dynamics~\cite{Rouzegar2022, Gorchon2022}. However, very recent experimental~\cite{Gorchon2022} and phenomenological analyses~\cite{Schmidt2023} find that assuming a straightforward connection between outgoing THz signal, the laser pulse and the spin physics can be highly misleading. 
	These issues could be resolved rigorously by a microscopic 
	and first-principles approach, as offered by TDDFT+Maxwell scheme~\cite{Noda2019, Tancogne-Dejean2020} where TDDFT calculations are coupled in multiscale and self-consistent  fashion to the Maxwell equations. But such a task has only recently been initiated in TDDFT software development~\cite{Noda2019, Tancogne-Dejean2020}, and its application to potentially large supercells of magnetic multilayers and necessary usage of noncollinear exchange-correlation (XC) functional~\cite{Eich2013a,Egidi2017}  is computationally very expensive. 

	\begin{figure*}
		\centering
		\includegraphics[width = \linewidth]{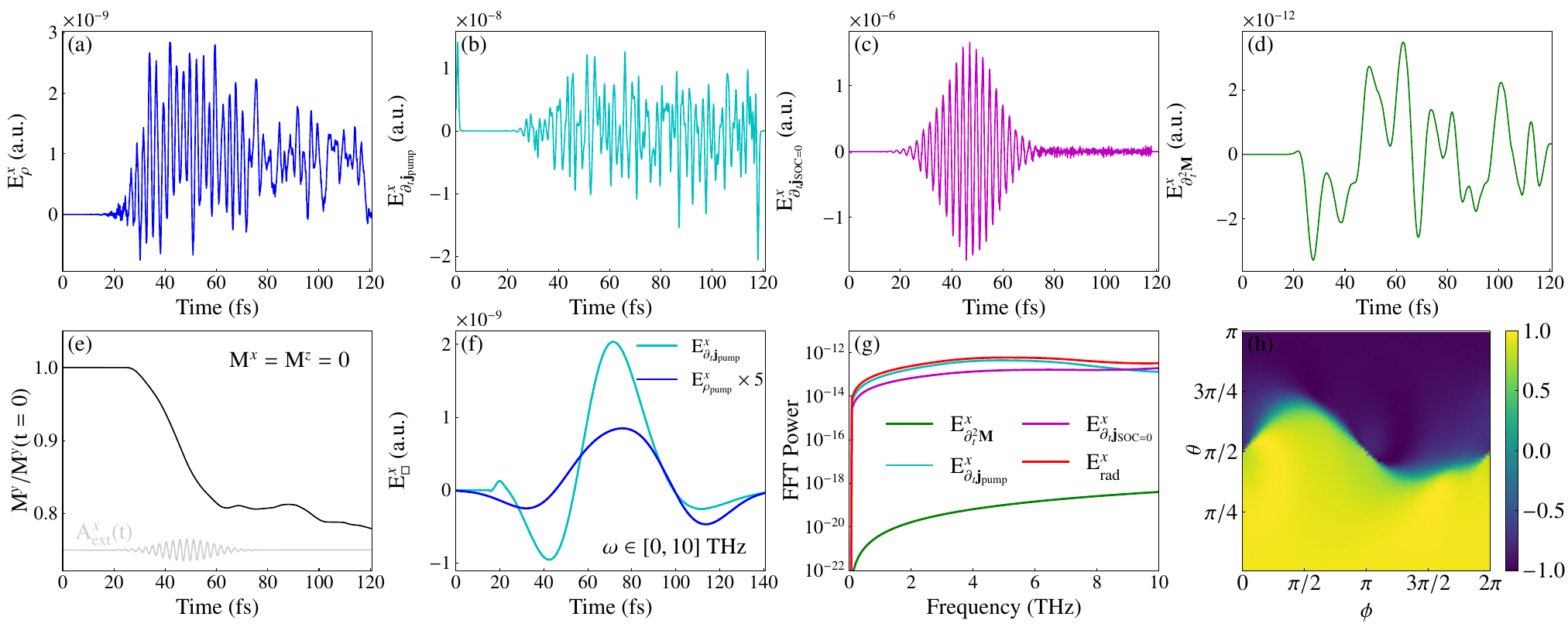}
		\caption{(a)--(d) Time-dependence of the $x$-component of $\mathbf{E}_\rho$ [Eq.~\eqref{eq:efieldrad}] as near-field, and three components [Eq.~\eqref{eq:eradni}] of far-field, EM radiation emitted by a single Ni layer driven by fs laser pulse [gray curve in (e) depicts its vector potential $A^x_\mathrm{ext}(t)$ in Eq.~\eqref{eq:TDDFT}]. (e) Demagnetization dynamics of Ni. (f) 
			Time-domain THz signal, radiated by charge (in near-field) or current (in far-field) density pumped by  demagnetization dynamics, as reconstructed from FFT components in (g) at frequencies \mbox{$\omega \in [0,10]$ THz}. 
			(g) FFT power spectrum of curves in (b)--(d), as well as of  their sum. (h) The angular distribution of (normalized) output energy $[E_\mathrm{out}^{\theta,\phi}]_\mathrm{norm}$, as obtained from  the  Poynting vector,  within the time interval of nonzero EM radiation emitted into the  solid angle $d\Omega$ at a distance 1000$a_B$.}
		\label{fig:fig2}
	\end{figure*}

\begin{figure*}
		\centering
		\includegraphics[width = \linewidth]{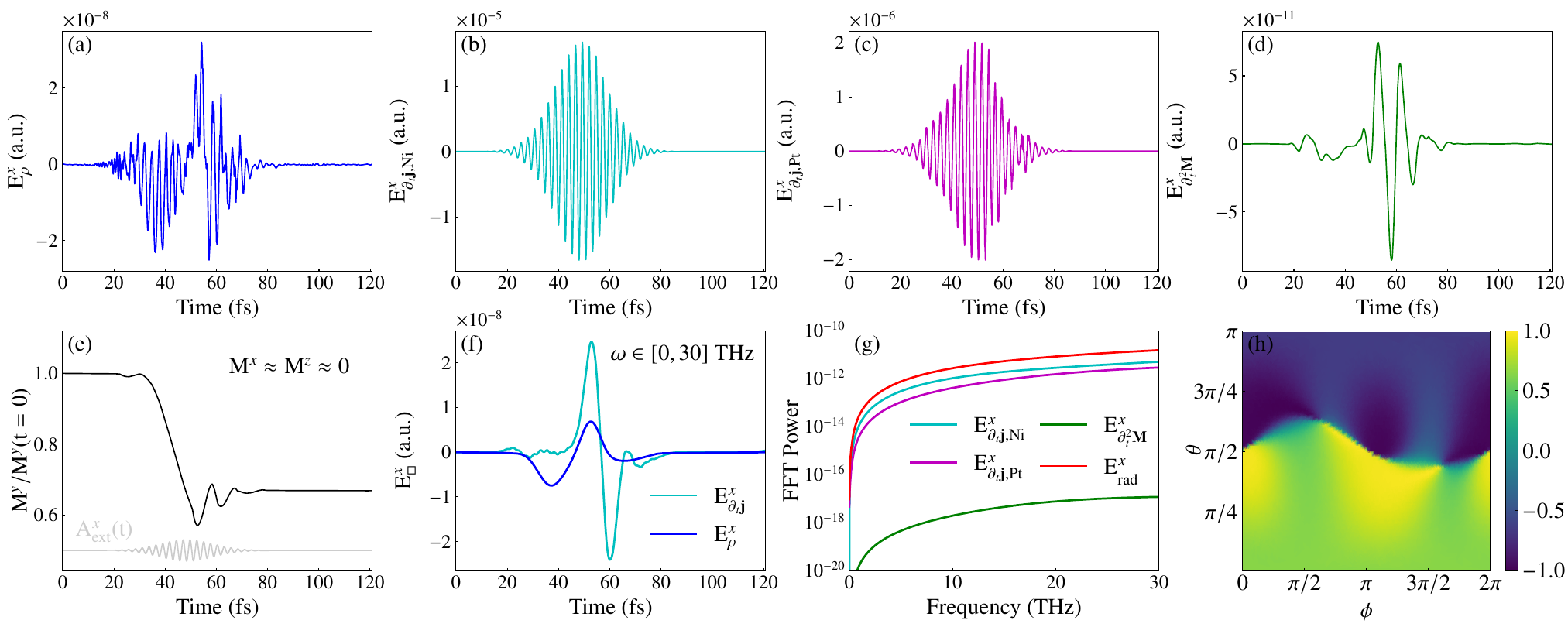}
		\caption{(a) Time-dependence of the $x$-component of $\mathbf{E}_\rho$ [Eq.~\eqref{eq:efieldrad}] as near-field, and  three components [Eq.~\eqref{eq:eradnipt}] of far-field, EM radiation emitted by  Ni/Pt bilayer driven by fs laser pulse [gray curve in (e) depicts its vector potential $A^x_\mathrm{ext}(t)$ in Eq.~\eqref{eq:TDDFT}]. (e) Demagnetization dynamics within Ni side of Ni/Pt bilayer. (f) 
			Time-domain THz signal, radiated by charge (in near-field) or current (in far-field) density, as reconstructed from FFT components in (g) at frequencies \mbox{$\omega \in [0,30]$ THz}. (g) FFT power spectrum of curves in (b)--(d), as well as of  their sum. (h) The same information as in Fig.~\ref{fig:fig2}(h), but for Ni/Pt bilayer.}
		\label{fig:fig3}
	\end{figure*}

	In this Letter, we introduce TDDFT+Jefimenko approach for computing  EM radiation from systems simulated by TDDFT, which is computationally inexpensive while offering direct insight into how time dependence of both uncompensated charge and current densities contribute to electric $\textbf{E}( \textbf{r},t)$ and magnetic $\textbf{B}( \textbf{r},t)$ fields of EM radiation. The insight is made possible by the structure of the Jefimenko formulas~\cite{Jefimenko1966}
	\begin{subequations}\label{eq:jefimenko}
		\begin{align}
			\textbf{E}(\textbf{r},t) & =  \frac{1}{4\pi \varepsilon_0} \int\!\! d^3r' \, \biggl[ \frac{\rho_\mathrm{ret}\textbf{R}}{R^3} + \frac{\partial_t \rho_\mathrm{ret} \textbf{R}}{cR^2}  -  \frac{ \partial_t \textbf{j}_\mathrm{ret}}{c^2R} 
			\biggl] \label{eq:efield}, \\
			\textbf{B}( \textbf{r},t) & =  \frac{\mu_0}{4\pi} \int\!\! d^3r' \, \biggl[ \frac{\textbf{j}_\mathrm{ret}}{R^3} + \frac{\partial_t \textbf{j}_\mathrm{ret} }{cR^2} \biggl] \times \textbf{R}, \label{eq:bfield}
		\end{align}
	\end{subequations}
	which can be viewed~\cite{Griffiths1991} as time-dependent generalizations of the Coulomb and Biot-Savart law, respectively. They are, in fact, an integral solution of the Maxwell equations in the approximations: fields vanish at infinity; their sources are confined to a finite region of space; and self-consistent effects, such as emitted EM radiation exerting backaction~\cite{Tancogne-Dejean2020,Philip2018} onto the source, can be neglected. They are properly retarded due to relativistic causality, i.e., the sources in Eq.~\eqref{eq:jefimenko} are  charge density $\rho_\mathrm{ret}(\textbf{r},t) = \rho(\textbf{r}, t-R/c)$ and charge current density $\textbf{j}_\mathrm{ret}(\textbf{r},t) = \textbf{j}(\textbf{r}, t-R/c)$, as well as their time derivatives, computed at the retarded time $t-R/c$.  Here $c$ is the velocity of light and $\mathbf{R}=\mathbf{r}-\mathbf{r'}$  is the vector from source at a point $\mathbf{r}'$ within magnetic heterostructure to the observation point $\mathbf{r}$. We also use shorthand notation $\partial_t \equiv \partial/\partial_t$ and $R=|\mathbf{R}|$. The observation  point is chosen as $\mathbf{r}=1000 a_B\mathbf{e}_z$, where $a_B$ is the Bohr radius, and the origin of the coordinate system is in the lower left corner of Ni layer [Fig.~\ref{fig:fig1}]. Since Jefimenko Eq.~\eqref{eq:jefimenko} operate with directly measurable quantities, they can provide clarity even for problems solvable using abstract EM potentials~\cite{Griffiths1991,Redzic2013}. For example, they show that $\partial_t \mathbf{j}$ is one of the sources of electric field [Eq.~\eqref{eq:efield}], while often invoked in THz spintronics literature  $\mathbf{E} \propto \mathbf{j}$---or $\mathbf{E} \propto \mathbf{j}_\mathrm{spin}$, assuming unwarrantedly~\cite{Schmidt2023,Gorchon2022} [see also Fig.~\ref{fig:fig3}(g)]  a one-to-one correspondence between spin $\mathbf{j}_\mathrm{spin}$ and charge currents---cannot be justified from Eq.~\eqref{eq:efield}. However, Jefimenko Eq.~\eqref{eq:efield} seems to suggest that EM radiation has both transverse and longitudinal electric field components decaying as $\propto 1/R$. This is only apparent as Eq.~\eqref{eq:efield} can be re-written~\cite{McDonald1997}, using the continuity equation, as
	\begin{eqnarray}\label{eq:efieldrad}
		\mathbf{E}(\mathbf{r},t) & = &  \frac{1}{4\pi \varepsilon_0} \int\!\! d^3r' \, \biggl[\, \underbrace{\frac{\rho_\mathrm{ret}\textbf{R}}{R^3}}_{\mathbf{E}_{\rho}} - \frac{\mathbf{j}_\mathrm{ret}}{cR^2}+ 2\mathbf{R} \frac{\mathbf{j}_\mathrm{ret} \cdot \mathbf{R} }{cR^4}  \nonumber \\
		&& \underbrace{\textbf{R} \frac{\partial_t \mathbf{j}_\mathrm{ret} \cdot \mathbf{R}}{c^2R^3}  - \frac{\partial_t \textbf{j}_\mathrm{ret}}{c^2R}}_{\mathbf{E}_{\partial_t \mathbf{j}}} \, \biggl].
	\end{eqnarray}
	Here, only the last two terms on the right-hand side (RHS), denoted as $E_{\partial_t \mathbf{j}}$, contribute to {\em far-field} EM radiation decaying as $\propto 1/R$. Rather than thinking in terms of microscopic sources of EM radiation, as divulged by Eq.~\eqref{eq:jefimenko}, analyses of  EM radiation in THz spintronics experiments have often employed plausible assumptions~\cite{Beaurepaire2004,Huang2019,Zhang2020}, such as that  a single FM layer emits magnetic dipole radiation in THz range whose electric field is given by~\cite{Griffiths2011} 
	\begin{equation}\label{eq:MagD}
		\mathrm{E}_{\partial_t^2 \textbf{M}}
		(\textbf{r},t) = \frac{1}{4\pi \varepsilon_0 c^3} \frac{\mathbf{R}}{R^2} \times \partial_t^2 \textbf{M}(\textbf{r}, t-R/c),
	\end{equation}
	where we use  $\partial_t^2 \equiv \partial^2/\partial t^2$. Since it decays as $\propto 1/R$, we include this contribution to the total electric field of far-field EM radiation, \mbox{$\textbf{E}_{\mathrm{rad}} = \mathbf{E}_{\partial_t \mathbf{j}} + \textbf{E}_{\partial_t^2 \textbf{M}}$}.
	
	
	{\em Results and Discussion.}---Our principal results are shown in Figs.~\ref{fig:fig2} and ~\ref{fig:fig3}. The charge response to a laser pulse is instantaneous [Figs.~\ref{fig:fig2}(a)--(c) and ~\ref{fig:fig3}(a)--(c)], while demagnetization [Figs.~\ref{fig:fig2}(e) and ~\ref{fig:fig3}(e)] lags behind the pulse. Magnetization shrinks along the $y$-axis, while the other two components are zero in Ni case or negligible in Ni/Pt case. We note that damagnetization features of Ni/Pt [Fig.~\ref{fig:fig3}(e)], with dip around \mbox{$\simeq 50$ fs} and slight recovery around \mbox{$\simeq 60$ fs}, are quite similar to experiments of Ref.~\cite{Siegrist2019} on the same system. Since the nonequilibrium charge dislocated into interatomic  space~\cite{Pellegrini2022} cannot be compensated by the positive background charge of ionic cores, it generates $\rho$ and $E^x_\mathrm{\rho}$ [Figs.~\ref{fig:fig2}(a) and ~\ref{fig:fig3}(a)] contributing to {\em near-field} EM radiation $E^x_{\rho} \propto 1/R^2$. This is relevant only at distances of the order of wavelength from the source, which is a few $\mu$m in the case of THz frequencies. The Fourier transform (FFT) of $E^x_{\rho}$ does have nonzero power at THz frequencies, even for NM layer alone, but since the near-field region is not scanned experimentally~\cite{Beaurepaire2004,Huang2019,Rouzegar2022,Seifert2016,Wu2017}, no THz radiation is detected from laser pulse-driven NM layers~\cite{Wu2017}. 
	
	In prior~\cite{Krieger2015} TDDFT simulations,  demagnetization was  explained as a two-step process involving two types of electrons---the valence ones respond fast to laser pulse by going into higher-energy delocalized states, with subsequent reduction and change in orbital angular momentum of atoms leading also to spin-flip, due to SOC, and the ensuing demagnetization on longer timescale of electrons closer to atomic cores where the strength of SOC is larger. However, there is still no consensus on indispensable microscopic mechanisms behind  ultrafast demagnetization. For example, laser pulse-driven spin-polarized~\cite{Battiato2010,Rudolf2012} currents, or even unpolarized one~\cite{Eschenlohr2013,Vodungbo2016} (when laser pulse exciting NM caping layer, so that photons never reach FM layer), of hot electrons have been invoked as alternative mechanisms to spin-flips~\cite{Rudolf2012,Eschenlohr2013,Vodungbo2016}. It is clear, however, that SOC$\neq$0 [forth term on the RHS of Eq.~\eqref{eq:TDDFT}] is {\em sine qua non} to obtain 
	demagnetization in TDDFT simulations~\cite{Krieger2015,Elliott2022}. Thus, the case SOC$=$0 for which magnetization is constant, $M^y(t)/M^y(t=0)=1$, appears uninteresting and was not analyzed in prior TDDFT studies. 
	Nevertheless, electrons of FM layer with SOC$=$0 will also respond to laser pulse, as is the case of any nonmagnetic layer~\cite{Vodungbo2016}, so that  examining the difference in charge current in 
	SOC$\neq$0 vs. SOC$=$0 cases can reveal  whether {\em demagnetization dynamics itself can generate charge and spin currents, on the top of the ones driven by laser pulse}. For example, several experiments have been interpreted~\cite{Rouzegar2022,Lichtenberg2022} as the consequence of  ``$dM/dt$ mechanism''~\cite{Choi2014} where pumped spin current $\propto \partial_t M_y$ is conjectured to exist and then converted by adjacent NM layer into charge current. However, in the standard model~\cite{Tserkovnyak2005} and its extensions ~\cite{Mahfouzi2012,Dolui2020b,VarelaManjarres2023} of spin current pumping by magnetization dynamics {\em only} rotation of its vector of {\em fixed} length is allowed, i.e., precession, so that spin current~\cite{Tserkovnyak2005} $\propto \mathbf{M} \times d\mathbf{M}/dt$ is apparently {\em zero} when $\mathbf{M}(t)$ is shrinking in length while not rotating [Figs.~\ref{fig:fig2}(e) and ~\ref{fig:fig3}(e)]. We also note that direct pumping of charge current by magnetization precession, without any spin-to-charge conversion necessary, is rarely found as it requires a number of conditions to be satisfied~\cite{Mahfouzi2012,Chen2009,VarelaManjarres2023,Ciccarelli2014,Hals2010}.
	
	Here we focus on possibility of direct pumping of charge current by demagnetization dynamics, extracted as  \mbox{$\mathbf{j}_\mathrm{pump}=\mathbf{j}_{\mathrm{SOC} \neq 0} - \mathbf{j}_{\mathrm{SOC}=0}$}~\footnote{See Supplemental Material at \url{https://wiki.physics.udel.edu/qttg/Publications} for time-dependence of total charge currents $I^\alpha_\mathrm{pump}(t)$, $I^\alpha(t)$, $I^\alpha_\mathrm{Ni}(t)$ and $I^\alpha_\mathrm{Pt}(t)$, obtained by integrating the respective charge current densities $j^\alpha_\Box(\mathbf{r},t)$ over the volume of Ni layer or Pt layer  or Ni/Pt bilayer [as denoted by the subscript within $\Box$ of $I^\alpha_\Box$ or $j^\alpha_\Box(\mathbf{r},t)$].} and we also define \mbox{$\rho_\mathrm{pump}=\rho_{\mathrm{SOC} \neq 0} - \rho_{\mathrm{SOC}=0}$}. We find \mbox{$j^x_{\mathrm{SOC}=0} \neq  0$}~\footnotemark[1] solely due to laser pulse, with thereby generated $E^x_{\partial_t j^x_{\mathrm{SOC}=0}}$  of far-field EM radiation shown in Fig.~\ref{fig:fig2}(c).  Therefore,  \mbox{$\mathbf{j}_\mathrm{pump}=(j^x_\mathrm{pump}, j^y_\mathrm{pump}, j^z_\mathrm{pump})$} quantifies excess charge current pumped by demagnetization dynamics, \mbox{$M^y(t)/M^y(t=0)<1$}, with the corresponding $E^x_{\partial_t j^x_\mathrm{pump}}$ shown in Fig.~\ref{fig:fig2}(b). In addition, $j^z_{\mathrm{SOC}=0} \equiv 0$~\footnotemark[1] means that {\em no}  charge (or spin) current flows along the $z$-axis if \mbox{$\mathbf{M}(t)=(0,\mathrm{const.},0)$}, so that  $j^z_\mathrm{pump} \equiv j^z_{\mathrm{SOC} \neq 0}$~\footnotemark[1] is solely due to demagnetization dynamics. Correspondingly, we break electric field of far-field EM radiation of single FM layer as
	\begin{equation}\label{eq:eradni}
		\textbf{E}_{\mathrm{rad}} = \mathbf{E}_{\partial_t \mathbf{j}_\mathrm{pump}} + \mathbf{E}_{\partial_t \mathbf{j}_\mathrm{SOC=0}} +  \textbf{E}_{\partial_t^2 \textbf{M}}. 
	\end{equation}
	The FFT of these  contributions reveals [Fig.~\ref{fig:fig2}(g)] that in THz range probed experimentally~\cite{Rouzegar2022,Seifert2016,Wu2017},  $\textbf{E}_{\partial_t^2 \textbf{M}}$ contribution [Eq.~\eqref{eq:MagD}] often invoked intuitively~\cite{Beaurepaire2004,Huang2019,Rouzegar2022} is {\em negligible} [Figs.~\ref{fig:fig2}(d) and ~\ref{fig:fig2}(g)] because of being $1/c$ smaller 
	than $\mathbf{E}_{\partial_t \mathbf{j}}$. In the same frequency range, $\textbf{E}_{\mathrm{rad}}$ is dominated by $\mathbf{E}_{\partial_t \mathbf{j}_\mathrm{pump}}$, thereby demonstrating that {\em charge current pumping by demagnetization is the key, but overlooked}, mechanism behind measured~\cite{Beaurepaire2004,Huang2019,Rouzegar2022} THz radiation of single ultrafast-light-driven FM layer. 
	
	In the case of Ni/Pt bilayer, the near-field EM radiation  governed by $\mathbf{E}_\rho$ [Eq.~\eqref{eq:efieldrad}] increases [Figs.~\ref{fig:fig3}(a) and ~\ref{fig:fig3}(f)] when compared to single Ni layer [Figs.~\ref{fig:fig2}(a) and ~\ref{fig:fig2}(f)]. We split far-field EM radiation from Ni/Pt bilayer
	\begin{equation}\label{eq:eradnipt}
		\textbf{E}_{\mathrm{rad}} = \mathbf{E}_{\partial_t \mathbf{j},\mathrm{Ni}} + \mathbf{E}_{\partial_t \mathbf{j},\mathrm{Pt}} +  \textbf{E}_{\partial_t^2 \textbf{M}}, 
	\end{equation}
	to find $E^x_\mathrm{rad}$, surprisingly, being dominated by radiation from charge current $j^x_\mathrm{Ni}$~\footnotemark[1] within Ni side [compared Fig.~\ref{fig:fig3}(b) vs. ~\ref{fig:fig3}(c), or cyan vs. magenta curves in Fig.~\ref{fig:fig3}(g)]. Thus, our results transform  standard picture~\cite{Rouzegar2022,Seifert2016,Wu2017}
	where $E^x_\mathrm{rad} \simeq E^x_{\partial_t \mathbf{j},\mathrm{Pt}}$ is  expected due to charge current $j^x_\mathrm{Pt}$~\footnotemark[1] in Pt generated by ISHE-based conversion of interlayer [i.e., flowing along the $z$-axis in the coordinate system of Fig.~\ref{fig:fig1}] spin current of {\em unclear} microscopic origin~\cite{Rouzegar2022}. In fact, its origin is the same as in the case of charge current within Ni layer---it is pumped by demagnetization dynamics. Figures~\ref{fig:fig2} and ~\ref{fig:fig3} clarify that charge current along the $x$-axis responsible for far-field EM radiation, in either FM or FM/NM systems, is present even if spin-to-charge conversion via SOC-based mechanisms~\cite{Saitoh2006,Jungfleisch2018a, Gueckstock2021,Wang2023} is  nonexistent (as in single FM layer) or inefficient. Our results are also fully compatible with recent experiments~\cite{Rouzegar2022,Huang2019} finding that THz radiation from both FM and FM/NM systems is directly related to 
	$\partial_t M^y(t)$, where the presence of NM layer in our picture simply accelerates demagnetization rate and increases  charge current pumping by demagnetization dynamics. That is, $M^y(t)$ in Fig.~\ref{fig:fig3}(e) is faster and with deeper reduction than in Fig.~\ref{fig:fig2}(e), as observed in numerous experiments~\cite{Malinowski2008,Kuiper2014}. This effect  can be attributed  to both the presence of FM/NM interface~\cite{Krieger2017} and to 
	SOC proximity effect~\cite{Dolui2020b,MarmolejoTejada2017} by which SOC within FM, and thereby any spin-flip scattering mechanism dependent on it~\cite{Krieger2017},  is strongly modified by hybridization of FM and NM wavefunctions~\cite{MarmolejoTejada2017,Dolui2020b}. Finally, by reconstructing the time-domain THz signal, from only the Fourier components at THz frequencies, we obtain curves in Figs.~\ref{fig:fig2}(f) and \ref{fig:fig3}(f) that are remarkably similar to typical THz signal reported  experimentally~\cite{Rouzegar2022,Seifert2016,Wu2017}. 
	
	We also examine the Poynting vector, \mbox{$\textbf{S}(\textbf{r},t) = \frac{1}{\mu_0}\textbf{E}(\textbf{r},t) \times \textbf{B}(\textbf{r},t)$}, of incoming light and outgoing EM radiation to compute the efficiency of conversion, $|E_{\rm out}|/|E_{\rm in}|$, which is \mbox{$7.66\times 10^{-5}$} for Ni layer and \mbox{$3.50\times 10^{-4}$} for Ni/Pt bilayer. The power radiated into the solid angle \mbox{$d\Omega=\sin \theta d\theta d\phi$} at distance $R$ in the direction $\mathbf{R}/R$ is given by  \mbox{$dP=\textbf{S}\cdot \mathbf{R} R d\Omega$}. Integrating $dP$ over the time interval $[t_i,t_f]$ of nonzero emitted EM radiation gives total  output energy, \mbox{$E_\mathrm{out}^{\theta,\phi}=\int_{t_i}^{t_f}dt\, dP$}, as a function of the azimuthal ($\phi$) and polar ($\theta$) angles. The normalized output energy [Figs.~\ref{fig:fig2}(h) and ~\ref{fig:fig3}(h)]  is then defined by \mbox{$[E_\mathrm{out}^{\theta,\phi}]_\mathrm{norm} = (E_\mathrm{out}^{\theta,\phi} - [E_\mathrm{out}^{\theta,\phi}]_\mathrm{mean})/2\sigma_{E_\mathrm{out}^{\theta,\phi}}$}, where $[E_{\rm out}^{\theta,\phi}]_\mathrm{mean}$ is the average value of $E_{\rm out}^{\theta,\phi}$ within the interval $[t_i,t_f]$  and $\sigma_{E_{\rm out}^{\theta,\phi}}$ is its standard deviation. The total input energy $E_{\rm in}$ is constructed in the same fashion for the laser pulse irradiating Ni surface.

	
	
	
	{\em Models and Methods}. For the purpose of demonstrating the capability of the TDDFT+Jefimenko approach, we use a single layer of Ni or Ni/Pt bilayer driven by fs laser pulse [Fig.~\ref{fig:fig1}].  The thicknesses of Ni and Pt layers are three and two monolayers (MLs) along [100] direction, respectively. In the TDDFT formalism, the time-dependent Kohn-Sham (KS) equation for Pauli spinors of KS orbitals is given by (using $\hbar=1$)
	\begin{eqnarray}\label{eq:TDDFT}
		i\frac{\partial \psi_j(\textbf{r},t)}{\partial t} & = & \bigg [ \frac{1}{2} \bigg(-i\nabla+\frac{1}{c}\mathbf{A}_\mathrm{ext}(t) \bigg)^2 + v_s(\textbf{r},t) \nonumber\\
		& +& \frac{1}{2c} {\bm \sigma} \cdot \textbf{B}_s(\textbf{r},t)\nonumber\\
		& + &\frac{1}{4c^2}{\bm \sigma} \cdot \big( \nabla v_s(\textbf{r},t)\times -i\nabla\big) \bigg]\psi_j(\textbf{r},t),
	\end{eqnarray}
	where $\mathbf{A}_\mathrm{ext}(t)$ is the vector potential of the applied laser pulse; $v_s(\textbf{r},t) = v_{\mathrm{ext}}(\textbf{r},t) + v_{\mathrm{H}}(\textbf{r},t) + v_{\mathrm{XC}}(\textbf{r},t) $ is the effective KS potential, as the sum of the external potential $v_{\mathrm{ext}}$ provided by the nuclei (treated as point particles), the Hartree potential $v_{\mathrm{H}}$ and XC potential $v_\mathrm{XC}$;   $\mathbf{B}_s(\textbf{r},t)=\mathbf{B}_\mathrm{ext}(t)+\mathbf{B}_\mathrm{XC}(t)$ is the KS magnetic field with $\mathbf{B}_\mathrm{ext}$ being the external magnetic field and $\mathbf{B}_\mathrm{XC}$ the XC magnetic field; \mbox{${\bm \sigma}=(\sigma_x, \sigma_y, \sigma_z)$} is the vector of the Pauli matrices; and the last term on the RHS describes SOC which necessitates usage of noncollinear XC functionals~\cite{Eich2013a,Egidi2017} even when long-range noncollinearity of local magnetization does not play a significant role. The  particle density of an interacting electronic system, as the fundamental quantity in (TD)DFT, is obtained from noninteracting KS electrons orbitals through \mbox{$n(\mathbf{r},t)=\sum_j \psi_j^\dagger(\mathbf{r},t) \psi_j(\mathbf{r},t)$}. Similarly, magnetization density, as an additional fundamental quantity in noncollinear (TD)DFT, is obtained from \mbox{$\mathbf{m}(\mathbf{r},t)=\sum_j \psi_j^\dagger(\mathbf{r},t) {\bm \sigma} \psi_j(\mathbf{r},t)$}, so that total magnetization is given by \mbox{$\mathbf{M}(t)=\int\!\! d^3 r\, \mathbf{m}(\mathbf{r},t)$}. Since the system is electrically neutral in equilibrium due to positive background charge of ionic cores, we subtract equilibrium charge density $en(\mathrm{r},t=0)$ from the nonequilibrium one $en(\mathrm{r},t)$ in order to obtain charge,  \mbox{$\rho(\mathbf{r},t-R/c)=en(\mathrm{r},t-R/c) - en(\mathrm{r},t=0)$}, dislocated into interatomic space~\cite{Pellegrini2022} and plug it into Eq.~\eqref{eq:efield}. We employ adiabatic local density approximation (ALDA) for XC functional~\cite{Lacombe2023} within full-potential linearized augmented plane-wave method as implemented in the ELK code~\cite{Dewhurst2016,elk}. The GS is also obtained from ELK using noncollinear static DFT calculations with  LDA XC functional. The grid of $\mathbf{k}$ vectors is chosen as $7 \times 7$ for Ni and $5 \times 5$ for Ni/Pt. After obtaining the GS, the dynamics for TDDFT calculations is generated by applying a Gaussian laser pulse with: central wavelength \mbox{$800$ nm};  $\simeq 50$ fs pulse duration; the peak intensity of  $5.45$ TW/cm$^2$; and 272.5 mJ/cm$^2$ fluence. Since the wavelength of applied laser light is much larger than the supercell, we assume dipole approximation and disregard spatial dependence of the vector potential $\mathbf{A}_\mathrm{ext}(t)$.

	
	
	{\em Conclusions and Outlook.}---In conclusion, despite being one of the principal  measured~\cite{Beaurepaire2004,Huang2019,Rouzegar2022,Seifert2016,Zhang2020,Wu2017}   quantities in experiments on ultrafast demagnetization, emitted THz  radiation is rarely calculated~\cite{Suresh2023,Nenno2019}, and never in prior TDDFT~\cite{Krieger2015, Krieger2017, Dewhurst2018, Dewhurst2021, Chen2019a, Chen2019c, Mrudul2023, Simoni2017, Stamenova2016,Acharya2020,Elliott2022,Pellegrini2022,Simoni2022} simulations which have provided some of the most detailed microscopic insights. Our coupling of time-dependent charges and currents from TDDFT to the Jefimenko equations demonstrates a first-principles route to obtain EM radiation, thereby allowing 
	us to replace guesses about the origin of charge current [Eq.~\eqref{eq:efieldrad}] and far-field EM radiation by it with microscopic and rigorous picture---charge current in Ni layer, as well as major contribution to it in Ni/Pt bilayer, originates from {\em previously unexplored}  phenomenon of charge pumping by magnetization changing length while not rotating. The present  TDDFT simulations~\cite{Krieger2015, Krieger2017, Dewhurst2018, Dewhurst2021, Chen2019a, Chen2019c, Mrudul2023, Simoni2017, Stamenova2016,Acharya2020,Elliott2022,Pellegrini2022,Simoni2022} are limited to \mbox{$\simeq 100$ fs} timescale and capture only 
	demagnetization~\cite{Tengdin2018} while harboring nonequilibrium charge within interatomic spaces which never relaxes~\cite{Pellegrini2022}.  On longer \mbox{$\gtrsim 500$ ps} timescales~\cite{Tengdin2018}, magnetization slowly recovers~\cite{Kuiper2014,Tengdin2018} while nonequilibrium charge density should relax \mbox{$\rho(\mathbf{r},t) \rightarrow 0$}~\cite{Pellegrini2022}, thereby also influencing current via the continuity equation, via electron-electron or electron-phonon scattering. Capturing such features would require TDDFT employing XC functionals beyond~\cite{Lacombe2023}  adiabatic ones and/or inclusion of dissipative bosonic environment (the latter direction is barely explored~\cite{Floss2019}). We relegate coupling of  such more complex TDDFT calculations to Jefimenko equations to future studies. 
	
\begin{acknowledgments}
	This research was supported by the US National Science Foundation (NSF) through the  University of Delaware Materials Research Science and Engineering Center, DMR-2011824. The supercomputing time was provided by DARWIN (Delaware Advanced Research Workforce and Innovation Network), which is supported by NSF Grant No. MRI-1919839.
\end{acknowledgments}

\end{document}